\documentstyle[12pt]{article}
\input{psfig}
\begin{document}
\vspace*{-6ex}
\begin{flushright}
\fbox{RU 95/E-06}
%\fbox{RU 95/E-xx} \fbox{ CDF/xxx/xxx/xxx/xxx}
\end{flushright}
\vglue 0.4cm
\begin{center}
{\Large\bf Pomeron flux renormalization\\
in soft and hard diffraction}\\
\vskip 1cm
K. GOULIANOS\\
{\em The Rockefeller University\\
1230 York Avenue, New York, NY 10021}\\
\vskip 1cm
Submitted to Physics Letters\\
February 14, 1995
%\today
\end{center}
\setlength{\baselineskip}{2.6ex}
\begin{center}
\parbox{13.0cm}
{\begin{center} ABSTRACT \end{center}
{\small \hspace*{0.3cm} While the main features of elastic, diffractive and 
total cross sections are described well by Regge theory, the 
measured  rise of the proton-(anti)proton 
single diffraction dissociation cross section with energy is 
considerably smaller than the theoretical prediction based on factorization 
and a constant triple-pomeron coupling.  
The observed energy dependence is obtained by renormalizing 
the pomeron flux ``carried" by a 
nucleon to unity.  Double diffraction and 
double pomeron exchange cross sections are reevaluated and compared to data, 
and a new interpretation of hard diffraction results  
emerges in which the hard pomeron obeys the momentum sum rule.}}
\end{center}
\parskip 12pt
\setlength{\baselineskip}{3ex}
\section{Introduction}
It is well known that pomeron exchange in Regge theory accounts for the 
main features of high energy 
elastic, diffractive and total cross sections \cite{KG,DL}.  
In particular, for proton-(anti)proton interactions, 
it accounts for the rise of the total cross section and
the shrinking of the forward elastic peak with energy, 
and also describes correctly 
the mass and $t$ dependence of single diffraction dissociation (SD).  
Furthermore, the concept of factorization provides relationships between 
cross sections that pass successfully the test of experimental observation 
\cite{KG}.  

Encouraged by this success, Ingelman and Schlein \cite{IS} 
proposed to extend factorization to the domain of hard processes 
involving the pomeron, and calculated the SD high $p_T$ dijet 
production cross section and rapidity distributions 
under various assumptions about a pomeron structure function.  In their 
calculation, they assumed that the pomeron has an independent existence 
inside a high energy proton (from now on we will use {\em proton} to refer 
to proton or antiproton) and defined a ``pomeron flux factor",
$f_{{\cal{P}}/p}(\xi,t)$,  
through the expression 
\begin{equation}
\frac{d^2\sigma_{SD}}{d\xi\,dt}=\sigma_T^{{\cal{P}}p}\;\; f_{{\cal{P}}/p}
(\xi,t)
\label{IS_flux}
\end{equation}
where $\sigma_{SD}$ is the SD cross section, $\xi$ is the fraction 
of the momentum of the proton carried by the pomeron, which is related to the 
pomeron-proton center of mass energy or diffractive mass $M$ by $\xi=M^2/s$, 
$t$ is the square of the four-momentum transfer or the negative mass 
squared of the (virtual) pomeron, and $\sigma_T^{{\cal{P}}p}$
is the pomeron-proton total 
cross section. Using a value for $\sigma_T^{{\cal{P}}p}$ 
obtained from the SD cross section at fixed target and ISR energies, 
the pomeron flux was evaluated at higher energies from 
the SD cross section through the above equation and was used to calculate 
hard pomeron-proton collisions in the usual way. 
When later 
the UA8 experiment studied diffractive dijet production, it was 
found that \cite{UA8}
while the shape of the 
rapidity distribution of the jets 
(we will use rapidity and pseudorapidity or $\eta$ interchangeably) 
favors an almost fully  
{\em hard} structure function for the pomeron, of the type 
$Q(\lambda)=6\lambda(1-\lambda)$, where $\lambda$ 
is the momentum fraction of a parton inside the pomeron, 
rather than a {\em soft} structure function of the type
$Q(\lambda)=6(1-\lambda)^5$, 
the rate of jet production 
can be accounted for by only a fraction of the momentum 
of the pomeron being carried by hard quarks or gluons. The UA8 rate result is
summarized by the ``discrepancy factor" required to multiply the pomeron
hard-quark or hard-gluon structure function
to predict the measured dijet rates. This 
factor is $0.46\pm0.08\pm0.24$  ($0.19\pm0.03\pm0.10$) for a hard-quark(gluon)
dominated pomeron \cite{UA8-EPS}.  

The discrepancy between the two UA8 results, 
namely the jet $\eta$ distribution requiring a hard pomeron structure function 
while the jet production 
rate being too small for a pomeron with a fully hard structure function, 
could be reconciled by 
noting that the pomeron, being a virtual state, need not observe the 
momentum sum rule \cite{UA8-EPS,DL88}. However, such a picture is not very 
satisfactory, as it puts into question the notion that the pomeron 
could have a structure function at all. Below, we show that this 
discrepancy can be attributed to the pomeron flux factor normalization, 
and that by {\em renormalizing} the flux to unity, i.e. to one pomeron per 
incident proton, the UA8 rate becomes consistent with the 
momentum sum rule. As a check of our 
flux normalization procedure, we show that the normalized flux 
predicts correctly the  energy dependence of the SD cross section using an 
energy independent triple-pomeron coupling.  In contrast, an unnormalized 
flux leads to a SD cross section which increases at a much faster rate 
than that observed experimentally, and predicts a rate
several times higher than the measured cross section at the 
Tevatron. 
\section{Pomeron flux factor}
It would appear that the flux factor defined by Eq.~\ref{IS_flux} 
could be obtained  
by dividing the SD differential cross section by $\sigma_T^{{\cal{P}}p}$.  
This is what was done by Ingelman and Schlein in their original calculation 
\cite{IS}, and more recently by Bruni and Ingelman in a calculation of the 
rates expected for diffractive $W$ production at the Tevatron \cite{BI}.  
The latter authors, for example, use a constant pomeron-proton total cross 
section of 2.3 mb and parameterize the flux factor as 
\begin{equation}
f_{{\cal P}/p}(\xi,t)=\frac{1}{2}\;\frac{1}{\xi}
\left[6.38\;e^{8t}+0.424\;e^{3t}\right]\frac{1}{2.3}
\label{BI_flux}
\end{equation}
This expression for the pomeron 
flux leads to a total integrated SD cross section of 9.1 mb for $\xi<0.05$  
at $\sqrt{s}$=546 GeV, in agreement with the value of $9.4\pm 0.7$ mb
reported by the UA4 experiment \cite{UA4} (the cross section is multiplied by 
a factor of 2 to account for the dissociation of {\em the other} nucleon).  

One problem with this approach is that it does not take into account the 
dependence  on $\xi$, expected in Regge theory, 
of the pomeron-proton total cross section and of the 
$t$ slope-parameter. Another problem 
is that the integral (for the standard diffractive region $\xi<0.1$)
of the flux factor over $t$ and $\xi$, which represents 
the total number of pomerons in the proton that participate in the diffractive 
interaction, grows with energy and has the value of 
2.2 (2.6) at $\sqrt{s}$=630 (1800) GeV. One may, therefore, ask the 
question: what does it really mean to have, 
say, two pomerons per proton?  How can {\em two} 
diffractive events be produced in {\em the same} $p \bar p$ collision?
Below we will show that this unnormalized flux leads to unphysical results, 
and that physically consistent results can be obtained only by normalizing 
the flux to unity. But first we discuss a flux factor 
that is consistent with Regge theory.

\begin{figure}[htb]
\psfig{figure=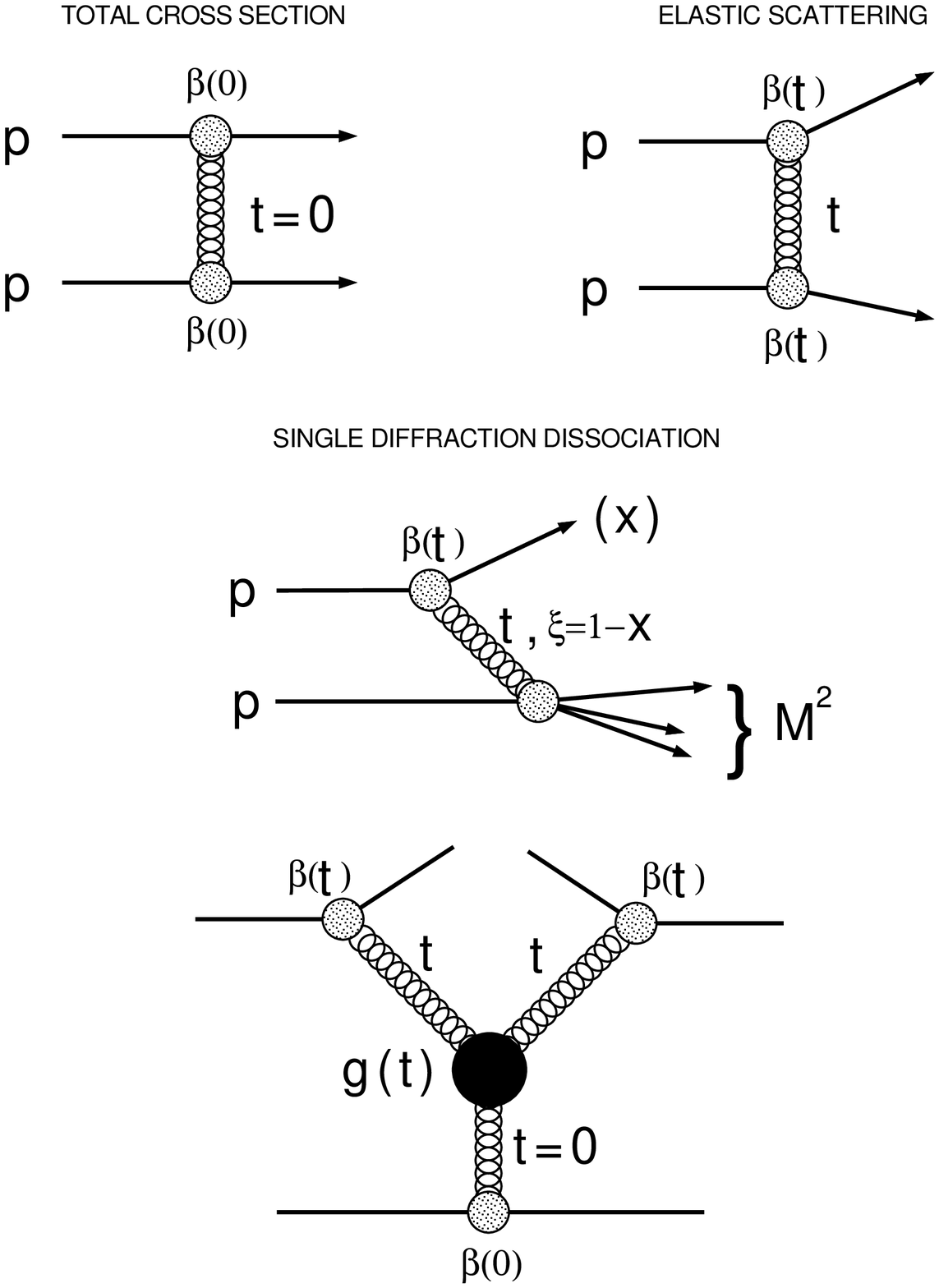,height=8in}
\caption{Feynman diagrams for the total, elastic, and single diffraction 
dissociation cross sections, including the ``triple-Regge" diagram for single 
diffraction.}
\label{diagrams}
\end{figure}

In terms of the pomeron trajectory,
\begin{equation}
\alpha(t)=1+\epsilon+\alpha't
\label{pomeron}
\end{equation}
the total, the elastic, and the SD $p\bar p$ cross sections can be written as
(see Fig.~\ref{diagrams})
\begin{equation}
\sigma_T=\beta^2(0)s^{\alpha(0)-1}=\sigma_0^{p\bar p}\; {s}^{\epsilon}
\label{total}
\end{equation}
\begin{equation}
\frac{d\sigma_{el}}{dt}=\frac{1}{16\pi(\hbar c)^2}\; 
\beta_1^2(t) \beta_2^2(t) \;
s^{2[\alpha(t)-1]}=\frac{\sigma_T^2}{16\pi(\hbar c)^2}\; 
e^{(2\alpha'ln{s})t}
\; F^4(t)
\approx \frac{\sigma_T^2}{16\pi(\hbar c)^2}\;e^{b_{el}(s)t}
\label{elastic}
\end{equation}
\begin{equation}
F^4(t)\approx e^{b_{0,el}t}\;\;\;
\Rightarrow\;\;\;b_{el}(s)=b_{0,el}+2\alpha'lns\;\;\;\;(s\;\;\mbox{in GeV}^2)
\label{slope}
\end{equation}
\begin{equation}
\frac{d^2\sigma_{SD}}{dtd\xi}=
\frac{1}{16\pi(\hbar c)^2 s}\;\beta_1^2(t)
\left(\frac{s}{M^2}\right)^{2\alpha(t)}
\beta_2(0)g(t)\left( M^2\right)^{\alpha(0)}=\frac{\sigma_0^{p\bar p}}
{16\pi(\hbar c)^2}\;\xi^{1-2\alpha(t)}\; F^2(t)\sigma_T^{{\cal{P}}p}
\label{diffractive}
\end{equation}
where $\beta(t)$ is the pomeron coupling to the proton, $g(t)$ the 
triple-pomeron coupling, $M$ the diffractive mass, F(t) the proton 
form factor, and $\sigma_T^{{\cal{P}}p}$  
the pomeron proton total cross section given by (see Eq.~\ref{total})
\begin{equation}
\sigma_T^{{\cal{P}}p}=\beta_1(0)g(t)\left( M^2\right)^{\alpha(0)-1}
=\sigma_0^{{\cal{P}}p}\left(M^2\right)^{\epsilon}
\label{Pp}
\end{equation}
where $M^2$ is in GeV$^2$, and 
in writing $\beta_1(0)g(t)=\sigma_0^{{\cal{P}}p}$ we have assumed that 
the triple-pomeron coupling constant, $g(t)$, is independent of $t$ \cite{KG}. 
From Eqs.~\ref{IS_flux} and \ref{diffractive} 
it is now clear that the flux factor can be 
expressed in terms of the total cross section, the elastic form factor, and the 
pomeron trajectory parameters as follows:
\begin{equation}
f_{{\cal{P}}/p}(\xi,t)=\frac{{d^2\sigma_{SD}}/{d\xi\,dt}}{\sigma_T^{{\cal{P}}p}}
=\frac{\sigma_0^{p\bar p}}
{16\pi(\hbar c)^2}\xi^{1-2\alpha(t)} F^2(t)
\label{R_flux}
\end{equation}

For the numerical evaluation of $f_{{\cal{P}}/p}(\xi,t)$ we use values
obtained from the recent
CDF results in Ref.~\cite{EDT}: $\sigma_T^{\bar{p}p}(s)=80.03\pm2.24$ mb at 
$\sqrt{s}$=1800 GeV,
$\epsilon=0.115\pm0.008$, and $\alpha'=0.26\pm0.02$.  The value of $\epsilon$ is 
the weighted average of three values: one obtained from the rise of
the total cross section with energy, $\epsilon=0.112\pm 0.013$, and 
the other two from 
the $\xi$-dependence of the SD
cross section at $\sqrt{s}=$546 GeV, $\epsilon=0.121\pm 0.011$, and at 
1800 GeV, 
$\epsilon=0.103\pm 0.017$. The value of $\alpha'$ is
obtained from a fit  to the form of 
Eq.~\ref{slope} of experimentally measured elastic scattering slope parameters 
at small-$t$ by CDF and at lower energies at the ISR (see \cite{EDT}).  
From the above values we obtain  
$\sigma_0^{p\bar p}=\sigma_T^{p\bar p}s^{-\epsilon}=14.3$ mb and 
$\frac{\sigma_0^{p\bar p}}
{16\pi(\hbar c)^2}=0.73\;GeV^{-2}$.  Note that 
at an energy as high as $\sqrt{s}$=1800 GeV the terms in the cross section 
that fall as $1/\sqrt{s}$ or faster are negligible and therefore 
Eq.~\ref{total} can be used directly to evaluate $\sigma_0^{p\bar p}$.  
The nucleon
form factor, $F(t)$, is obtained from elastic scattering.
In the small-t region, the t-dependence of elastic scattering is represented
well by $F^4(t)\approx e^{b_{0,el}t}$.
From the elastic slope parameter at $\sqrt{s}=$1800 GeV,
$b_{el}=16.98\pm0.25\; GeV^{-2}$ \cite{EDT}, using $\alpha'=0.26$
we obtain (Eq.~\ref{slope}) $b_{0,el}=b_{el}-2\alpha'lns=
9.2\;GeV^{-2}$ and hence $F^2(t)\approx e^{b_{0,SD}t}=
e^{(1/2)b_{0,el}t}=e^{4.6t}$.  This is consistent with the value
$b_{0,SD}=4.2\pm0.5\;GeV^{-2}$ measured in \cite{EDT} at $\sqrt{s}=$ 1800 GeV.
However, this  expression underestimates the
cross section at large t-values, and this is the reason for using two 
exponentials in Eq.~\ref{BI_flux}.  

Another expression for the flux factor consistent with Regge theory 
was proposed by Donnachie and Landshoff (DL) \cite{DL88}, who argue 
persuasively that the 
pomeron couples to quarks like an isoscalar photon and therefore 
the relevant form factor is $F_1(t)$, the
isoscalar form factor measured in electron-nucleon scattering 
\begin{equation}
F_1(t)=\frac{4m^2-2.8t}{4m^2-t}\left[\frac{1}{1-\frac{t}{0.7}}\right]^2
\label{F1}
\end{equation}
where $m$ is the mass of the nucleon.  The DL flux factor is given by
\begin{equation}
f_{{\cal{P}}/p}(\xi,t)=\frac{9\beta_0^2}{4\pi^2}\xi^{1-2\alpha(t)} F_1^2(t)
\label{DL_flux}
\end{equation}
where $\beta_0\approx1.8\;GeV^{-1}$ is the pomeron-quark coupling. With this 
value for $\beta_0$ we obtain $\frac{9\beta_0^2}{4\pi^2}=0.74$, which is 
to be compared with the 
value of $\frac{\sigma_0^{p\bar p}}{16\pi(\hbar c)^2}=0.73$ 
of Eq.~\ref{R_flux}. 
Since the discrepancy factor in the UA8 analysis is based on 
the DL form factor with a pomeron trajectory $\alpha(t)=1.08+0.25t$ 
\cite{UA8-EPS}, we will use  this trajectory and Eq.~\ref{DL_flux} 
when we recheck the momentum sum rule 
for the hard pomeron 
with a normalized pomeron flux.  However, in deriving the energy dependence 
of the SD cross section below, we will make use of the flux given by 
Eq.~\ref{R_flux} with a pomeron trajectory $\alpha(t)=1.115+0.26t$ and 
$F^2(t)=e^{4.6t}$, since it corresponds  to the expression used 
to derive the integrated SD cross sections from the (more accurate at high 
energies) CDF data.

\section{Single diffraction dissociation}  
The integral of the SD 
cross section can be written in terms of $M^2$ and $t$ as
(see Eqs.~\ref{IS_flux}, \ref{diffractive} \&\ref{Pp} and use $\xi=M^2/s$)
\begin{equation}
\sigma_{SD}
=C\;s^{2\epsilon}\;\sigma_0^{{\cal{P}}p}\;
\int_{t=0}^{t=\infty}\int_{M_0^2}^{M^2=0.1s}
\frac{(s/M^2)^{2\alpha't}}{\left(M^2\right)^{1+\epsilon}}\;
F^2(t){dtdM^2}
\label{SD}
\end{equation}
where $s$ and $M^2$ are in GeV$^2$, 
C=0.73, $\epsilon=0.115$,  
$\alpha'=0.26$,  $F(t)$ the proton form factor, and $M_0^2=1.4\;GeV^2$ is 
the {\em effective} diffractive threshold \cite{EDT}.
The only unknown parameter in this expression is 
$\sigma_0^{{\cal{P}}p}$, the pomeron-proton total cross section at $M^2=1$ 
GeV$^2$. This formula yields a ratio of the 
diffractive cross section at $\sqrt{s}$=546 to that at $\sqrt{s}$=20 GeV 
of 4.5, which is much larger than the experimental 
value of $\approx 1.6$ 
(see discussion on p. 5546 of Ref.~\cite{EDT}).  Clearly, the 
above expression does not give the correct energy dependence for 
$\sigma_{SD}$.  
The difference from experiment is almost entirely due to the factor 
$s^{2\epsilon}$, as pointed out in \cite{EDT}. The SD cross section 
as given by Eq.~\ref{SD} becomes larger than the total cross 
section at higher energies, violating unitarity. 

\begin{figure}[htb]
\vspace*{-1in}
\centerline{\psfig{figure=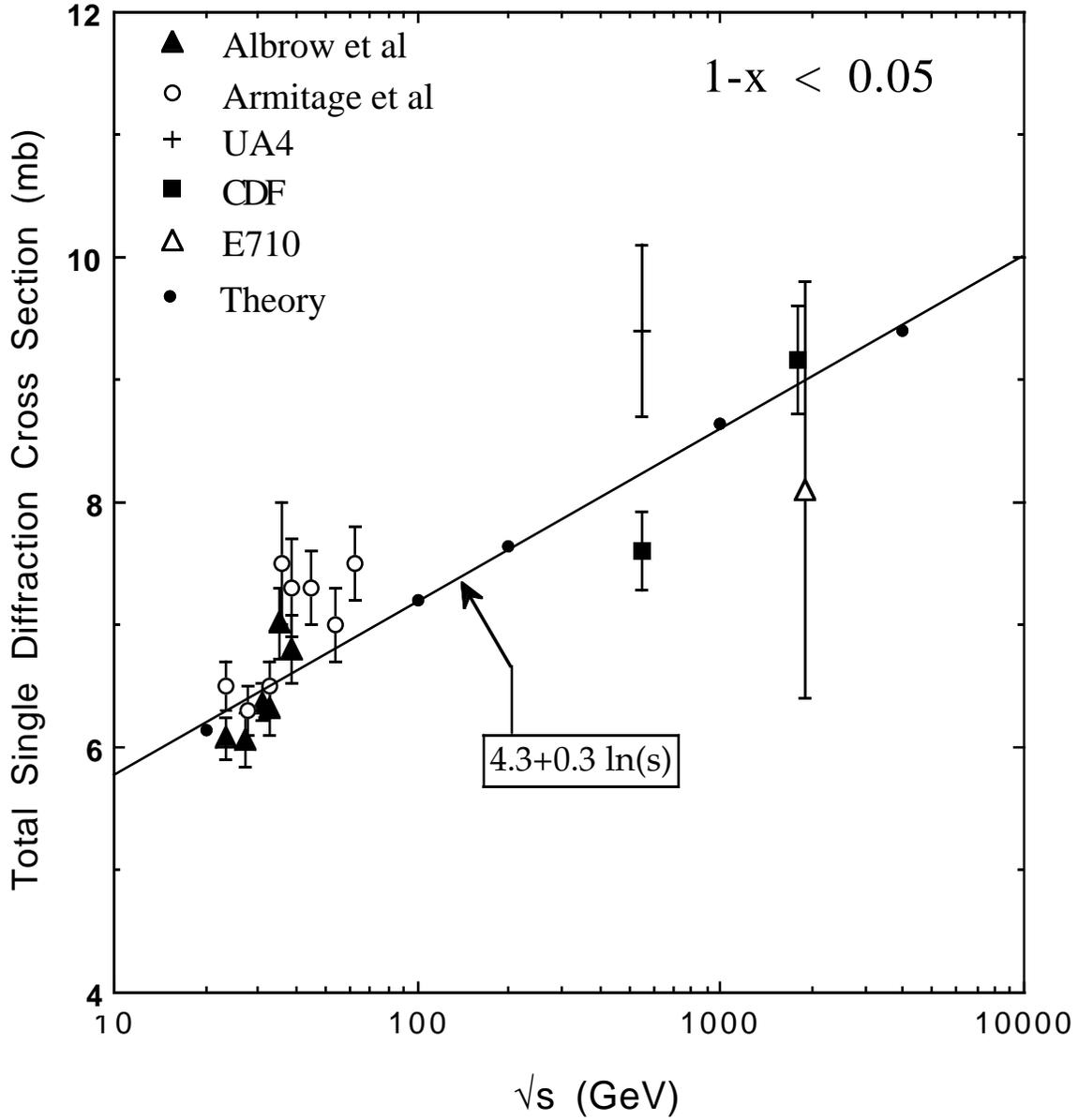,height=10in}}
\vspace*{-1in}
\caption{The total \protect{$p(\bar p)-p$} normalized
SD cross section  
for \protect{$\xi<0.05$}, calculated from \protect{Eq.~14} 
with \protect{$\sigma_0^{{\cal{P}}p}$=2.6 mb} and multiplied by 2,
is compared 
with experimental results.}
\label{fit}
\end{figure}

Let us now insist that no more than one pomeron per incident proton be 
allowed to participate in a diffractive process, i.e let us {\em re}-normalize 
the flux factor to unity:
\begin{equation}
f_s(\xi,t)\equiv \frac{f_{{\cal{P}}/p}(\xi,t)d\xi dt}{N(s)}=
\frac{f_{{\cal{P}}/p}(\xi,t)d\xi dt}
{\int_{M_0^2/s}^{\xi_{max}}\int_{t=0}^{\infty} f_{{\cal{P}}/p}(\xi,t)d\xi dt}
\label{FN}
\end{equation}
The integrated flux factor ($\xi_{max}=0.1$) of Eq.~\ref{R_flux} 
with $F^2(t)=e^{4.6t}$ is 
$N(s)=1$ at $\sqrt{s}=20$ GeV and increases with energy approximately 
as $s^{2\epsilon}$  reaching the value of 9.2 at $\sqrt{s}=1800$ GeV.  
The normalized integrated SD cross section is given by 
\begin{equation}
\sigma_{SD,N}=<\sigma_T^{{\cal{P}}p}>_{f_s}=\sigma_0^{{\cal{P}}p}\;
{s}^{\epsilon}\;
\int \xi^{\epsilon}f_s(\xi,t)d\xi dt
\label{SDN}
\end{equation}
where we have used Eq.~\ref{Pp}
with $\left( M^2\right)^{\epsilon}={s}^{\epsilon}\xi^{\epsilon}$.
Again, the only unknown in this equation is $\sigma_0^{{\cal{P}}p}$. 
The total single diffraction 
cross section calculated 
from Eq.~\ref{SDN} with $\sigma_0^{{\cal{P}}p}=$2.6 mb is compared 
in Fig.~\ref{fit} with experimental data from ISR \cite{ISR}, 
UA4 \cite{UA4}, E710 \cite{E710}, and CDF \cite{EDT}. For this particular 
comparison we 
used data and calculated cross sections for $\xi <0.05$ in order 
to reduce possible non-pomeron contributions to the data \cite{KG}.  
Considering the 
systematic uncertainties represented by the scatter in the data points, 
the agreement is good.  Without renormalizing the pomeron flux, the 
calculated cross section at $\sqrt{s}$=1800 GeV would be almost an order 
of magnitude higher!   From Eq.~\ref{SDN} 
it is clear that the normalized cross section rises 
with energy at a rate slower than ${s}^{\epsilon}$, staying 
safely below the total $p\bar p$ cross section, as required by unitarity.
An approximate expression for the rise of the total SD cross section with 
energy is given by (see Fig.~\ref{fit})
\begin{equation}
\sigma_{SD}^T=4.3+0.3\;lns\;\;\;\mbox{mb}\;\;\;(s\mbox{ in GeV}^2)
\end{equation}

\section{Pomeron-proton total cross section}

The pomeron-proton total cross section is related 
intimately to the normalized single diffraction cross section through 
Eq.~\ref{SDN}.  Fitting the data with this equation not only yields the 
constant $\sigma_0^{{\cal{P}}p}$ but also verifies the assumed 
$\sim (M^2)^{\epsilon}$ energy dependence, where $M^2=\mbox{\large \^{s}}$ 
is the 
pomeron-proton center of mass energy.  From this fit we therefore infer that 
\begin{equation}
\sigma_T^{{\cal{P}}p}=2.6\;\mbox{\large \^{s}}^{\epsilon}\;\;\mbox{mb}
\;\;\;(s\mbox{ in GeV}^2)
\end{equation}
where $\epsilon=0.115$ is the offset from unity of the 
intercept of the pomeron 
trajectory at $t=0$.  Thus, the pomeron behaves like a hadron. 
The ratio of $\sigma_0^{{\cal{P}}p}$ 
to $\sigma_0^{p\bar p}$ is 
\begin{equation}
\sigma_0^{{\cal{P}}p/{p\bar p}}=0.18
\label{rs}
\end{equation}
Since the uncertainty in the value of $\epsilon$ affects both the numerator and 
denominator of this ratio in approximately the same proportion, 
the value of the ratio is not sensitive to the error in $\epsilon$.  
The same is true for the ratio of the triple-pomeron to the pomeron-quark 
coupling constants discussed below.

\section{Triple-pomeron coupling constant}
 
From $\sigma_0^{{\cal{P}}p}$ we obtain the value of the 
triple-pomeron coupling constant (use Eqs.~\ref{Pp} \&\ref{total}), assuming 
that it is independent of $t$:
\begin{equation}
g(t)\equiv g(0)=
\frac{g(t)\beta(0)}{\beta^2(0)}=\frac{\sigma_0^{{\cal{P}}p}}
{(\sigma_0^{p\bar p})^{\frac{1}{2}}}=0.69\;{\mbox{mb}}^{\frac{1}{2}}=
1.1\;{\mbox{GeV}}^{-1}
\label{triple}
\end{equation}
This value of $g(t)$ is almost a factor of two higher than the value 
$g(t)=0.364\pm 0.025$ mb$^{\frac{1}{2}}$  reported in Ref.~\cite{E396}.  
This apparent discrepancy is due  
to the different parameterization ($\epsilon=0$ and
$\sigma_0^{p\bar p}=\sigma_T^{p\bar p}$) used in evaluating $g(t)$ 
from the data in \cite{E396}. 

If the pomeron couples to quarks, as proposed by DL \cite{DL88}, 
the pomeron-quark coupling constant may be evaluated by equating the 
coefficients of Eqs.~\ref{R_flux} \&\ref{DL_flux}, which yields
\begin{equation}
\beta_0=\frac{\sqrt{\pi\sigma_0}}{6(\hbar c)}=1.8\;\mbox{GeV}^{-1}
\label{quark}
\end{equation}
The ratio of the triple-pomeron to the pomeron-quark coupling, 
$g(t)$ to $\beta_0$, is given by
\begin{equation}
\frac{g(t)}{\beta_0}=0.61
\label{rc}
\end{equation}
Again, while the values of both $g(t)$ and $\beta_0$ are correlated with the 
value of $\epsilon$, their ratio is insensitive to the uncertainty in 
$\epsilon$.   

\section{Double diffraction dissociation}
\begin{figure}[htb]
\hspace*{-1cm}
\psfig{figure=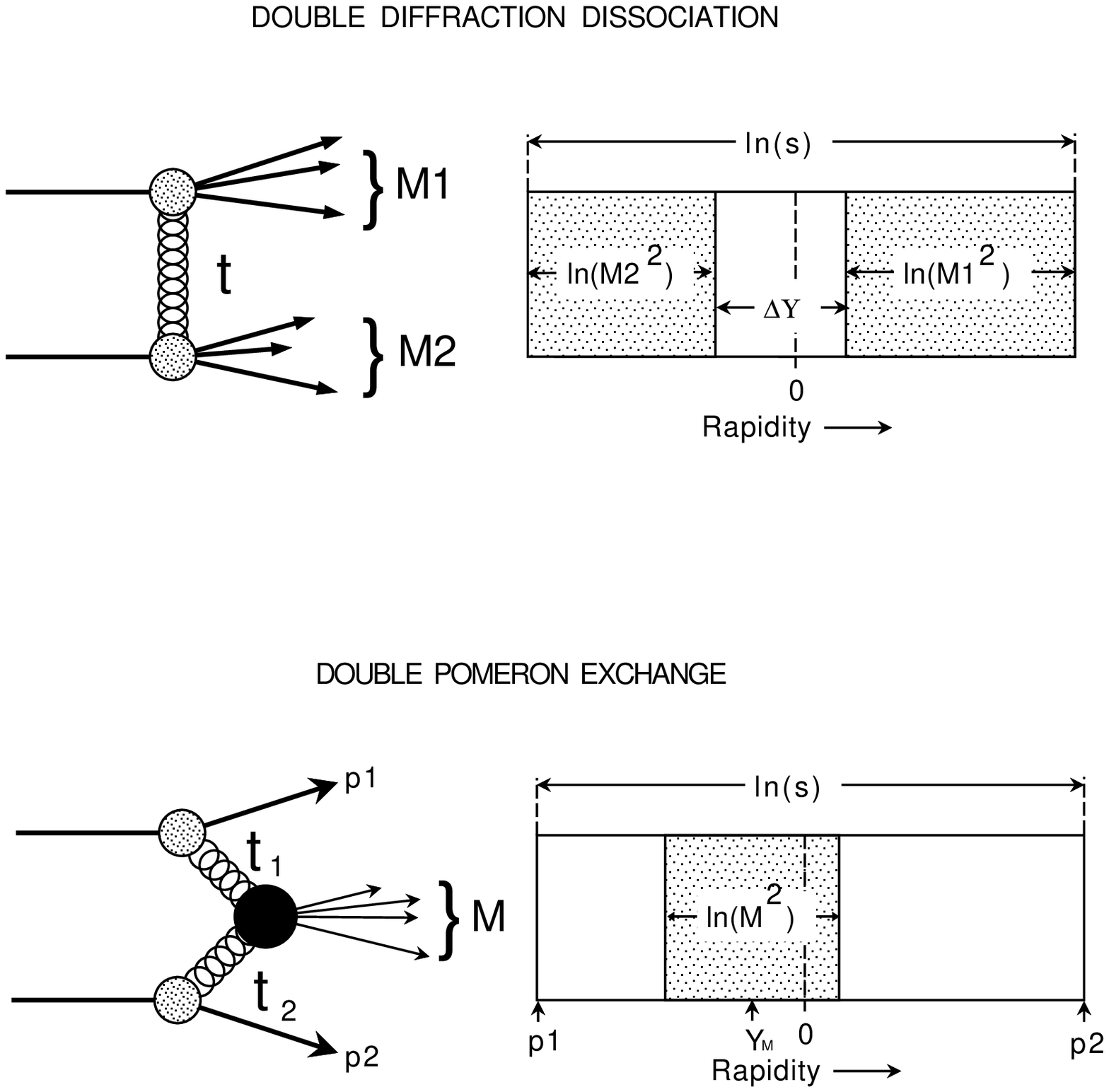,height=9in}
\vspace*{-1.5in}
\caption{Feynman diagrams, and rapidity regions occupied by the 
diffractive clusters, 
for double diffraction dissociation and for double pomeron exchange.}
\label{Fig_3}
\end{figure}

In double diffraction dissociation (DD) 
both nucleons dissociate, as shown in Fig.~\ref{Fig_3}. 
Assuming pomeron exchange and factorization, 
the DD cross section may be 
obtained from the SD and elastic scattering cross sections using 
Eqs.~\ref{IS_flux}, \ref{FN} \&\ref{elastic},
\begin{equation}
\frac{d^3\sigma_{DD}}{dM_1^2\,dM_2^2\,dt}=\frac{1}{d\sigma_{el}/dt}\;
\frac{d^2\sigma_1}{dM_1^2dt}\;\frac{d^2\sigma_2}{dM_2^2dt}
=\left(\frac{\sigma_0^{{\cal P}p}}{4\sqrt{\pi}\hbar c}\right)^2
\left(\frac{s^{\epsilon}}{N(s)}\right)^2
\frac{e^ { 2\alpha'ln\left[ \frac{s\;(1\,GeV^2)}{M_1^2M_2^2} \right] t} }
{(M_1^2M_2^2)^{1+\epsilon}}
\label{SDD}
\end{equation}
where  $N(s)$ 
is the integral of the pomeron flux factor (see Eq.~\ref{FN}).  
The nucleon form factor, $F(t)$, 
drops out in the division, so that the $t$-dependence is given by the 
slope parameter
\begin{equation}
b_{DD}=2\alpha' ln\left[\frac{s\;(1\,GeV^2)}{M_1^2M_2^2} \right]=
2\alpha'\,\Delta y \hspace*{0.7cm}\left[\mbox{GeV}^{-2}\right]
\end{equation}
where $\Delta y$ is the rapidity gap between the two diffractive clusters 
(see Fig.~\ref{Fig_3}).  
If we now apply the requirement $\Delta y>2.3$, 
which corresponds to the {\em coherence} requirement 
$(M^2/s)<0.1$ in single diffraction -- since $ln(s/M^2)>-ln(0.1)=2.3$, 
we obtain the {\bf coherence condition} for double diffraction:
\begin{equation}
\frac{M_1^2\,M_2^2}{s\;(1\,GeV^2)}<0.1
\label{coherence}
\end{equation}
With this condition as a constraint the $b_{DD}$ parameter is positive for 
all mass combinations.  If $\Delta y$ were to become negative, which would 
correspond to mass clusters overlapping in rapidity, $b_{DD}$ would 
become negative and the cross section would increase with $t$.  We therefore 
interpret Eq.~\ref{coherence} 
to mean that coherence breaks down for rapidity gaps 
smaller than $\sim 2.3$ units, and integrate Eq.~\ref{SDD} 
subject to the coherence condition 
to obtain the total DD cross section:
\begin{equation}
\sigma_{DD}=K(s)\int_{M_1^2=1.4}^{0.1s/1.4}\int_{M_2^2=1.4}^{0.1s/M_1^2}
\frac{dM_1^2\,dM_2^2}{(M_1^2\;M_2^2)^{1+\epsilon}\;ln(s/M_1^2\,M_2^2)}
\label{TDD}
\end{equation}
$$\mbox{where}\;\;\;\;
K(s)=\frac{1}{2\alpha'}
\left(\frac{\sigma_0^{{\cal P}p}}{4\sqrt{\pi}\hbar c}\right)^2
\left(\frac{s^{\epsilon}}{N(s)}\right)^2$$
Table~\ref{Table_1} lists cross sections at 
several energies calculated using this equation.  
The decrease of the cross section with energy is due to the 
faster increase of the elastic relative to the diffractive cross section. 
%*****************************
\begin{table}[htb]
\center
\caption{Total double diffraction cross sections.}
\vspace{1.0ex}
\begin{tabular}{|c|c|}
\hline
$\sqrt{s}\;\;[GeV]$&$\sigma_{DD}^T\;\; [mb]$\\
\hline
\hline
30&3.1\\
\hline
200&2.3\\
\hline
630&1.7\\
\hline
900&1.6\\
\hline
1800&1.3\\
\hline
14000&0.75\\
\hline
\end{tabular}
\label{Table_1}
\end{table}
%********************************

 A practical way of measuring the inclusive double diffractive cross section at 
hadron colliders is to look for events with a rapidity gap centered at
$y=0$.  Table~\ref{Table_2} lists the cross sections expected 
at the Tevatron, $\sqrt s=1800$ GeV, 
as a function of the width  $\Delta y$ of the rapidity gap. These cross 
sections were  calculated 
from Eq.~\ref{TDD} with $M^2_{1,max}=M^2_{2,max}=\sqrt{s}\;
e^{-\Delta y/2}$.  As shown, the 
cross section decreases slowly as the rapidity gap width increases.  

%*************************
\begin{table}[htb]
\center
\caption{$\sigma_{DD}$ versus $\Delta{y}$ at the Tevatron.}
\vspace{1.0ex}
\begin{tabular}{|c|c|}
\hline
$\Delta y$ (central)&$\sigma_{DD}^{\Delta y}\;[mb]$\\
\hline
\hline
2.0&0.62\\
\hline
2.5&0.57\\
\hline
3.0&0.52\\
\hline
3.5&0.47\\
\hline
4.0&0.42\\
\hline
4.5&0.39\\
\hline
\end{tabular}
\label{Table_2}
\end{table}
%**********************************

Using the rapidity gap technique, the UA5 collaboration measured the DD 
cross section at the CERN S$p\bar p$S collider and reported values 
of $3.5\pm 2.5\;(4.0\pm 2.2)$ mb at $\sqrt s=200 \;(900)$ GeV, respectively 
\cite{UA5}.  
These values are
within $1\;\sigma$ of those in Table~\ref{Table_1}, 
but are systematically higher. 
This  may be due to an underestimate of the detector acceptance for DD events, 
which was obtained with a Monte Carlo simulation where 
single diffractive clusters were generated on each side and 
were allowed to reach independently 
and simultaneously mass values up to $M^2_{max}=0.05s$.  This procedure 
allows overlapping diffractive clusters in violation of the coherence condition 
of Eq.~\ref{coherence}, resulting in a 
lower acceptance for DD events and hence a larger cross section.

At the Tevatron, where the energy of $\sqrt s$=1800 GeV provides a 
rapidity range of 15 units, accurate measurements of DD cross sections as a 
function of rapidity gap width can be performed using minimum bias data 
triggered by the ``beam-beam" counters.  Such data are 
already available in the CDF and D0 experiments.  The measurements can best be 
done by fitting the particle multiplicity distribution in a given region of
$\Delta \eta$ centered at $\eta=0$ and extracting from the fit the number 
of {\em excess} events in the zero multiplicity bin.  The fraction of these 
{\em rapidity gap} events to the total number of events in the sample can then 
be compared directly with the values in Table~\ref{Table_2} divided by 
the non-diffractive inelastic cross section of 50 mb.

\section{Double pomeron exchange}

In double pomeron exchange (DPE) two pomerons, 
one from each incoming hadron, interact to form a 
diffractive cluster of mass $M$ centered at rapidity $y_M$ 
(see Fig.~\ref{Fig_3}). 
The cross section for DPE is obtained from the SD and total cross sections 
using factorization (see \cite{KG2}):
\begin{equation}
\frac{d^4\sigma}{d\xi_1d\xi_2dt_1dt_2}=\frac{1}{\sigma_T^{p\bar p}}\;\;
\frac{d^2\sigma_1}{d\xi_1dt_1}\;\frac{d^2\sigma_2}{d\xi_2dt_2}
\end{equation}
The mass of the cluster and the rapidity of its centroid are related to the 
variables $\xi_{1,2}$:
\begin{equation}
M^2=s\,\xi_1 \xi_2
\end{equation}
$$y_M=\frac{1}{2}\,ln\frac{\xi_1}{\xi_2}$$
The condition $\xi_{1,2}<0.1$ for SD translates to the condition 
$$M^2<0.01\,s$$
Using Eqs.~\ref{total} \&\ref{IS_flux} 
with a normalized pomeron flux, and changing 
the variables from $\xi_{1,2}$ to $M^2$ and $y$, we obtain the expression 
\begin{equation}
\frac{d^2\sigma}{dM^2dy_M}=\sigma_0^{p\bar p}\left(\frac{\sigma_0^{{\cal{P}}p}}
{16\pi (\hbar c)^2}\frac{s^{\epsilon}}{N(s)}\right)^2\;
\left\{(M^2)^{1+\epsilon}\;\left[(b+\alpha'ln\frac{s}{M^2})^2-(2\alpha'y_M)^2
\right]\right\}^{-1}
\end{equation}
where $b=4.6$ is the slope parameter of the exponential proton form factor, 
$F(t)$, used here for simplicity. For a given mass $M$, $y_M$ 
varies within the range
$\pm \frac{1}{2}ln\frac{M^2}{0.01s}$, so that $|2y_M|<ln\frac{s}{M^2}$ and the 
term in the square brackets is a function decreasing logarithmically with 
increasing $M^2$.  As a result, the DP cross section falls approximately as 
$1/M^2$.  
A numerical integration of this equation for the range $1$ GeV$^2<M^2<0.01s$ 
yields an inclusive DPE 
cross section of 61, 76, 69 and 50 $\mu$b at $\sqrt s$= 50, 630, 1800 and 14000
GeV, respectively.  The calculated 
value of 76 $\mu$b at 630 GeV is in agreement 
with the experimental value of 30-150 $\mu$b reported by the UA8 experiment
\cite{UA8_DP}.   
The DP cross section is approximately constant 
through the entire range from the ISR to the LHC collider energies. 
On a finer scale, it rises initially with energy and then falls as the 
$lns$ term in the denominator becomes comparable to $b$.

\section{Hard diffraction}
According to our renormalization scheme, all rate predictions  
for hard processes in diffraction dissociation based on the
procedure suggested by Ingelman and Schlein  \cite{IS} 
must be scaled down by the integral of the pomeron flux factor at 
the given energy.  Since the flux factor is unity at $\sqrt{s}=20$ GeV and 
increases with energy as $\sim s^{2\epsilon}$, the scaling factor varies 
approximately as $(\sqrt{s}/20)^{4\epsilon}$.  This {\em renormalization} 
of the flux 
lowers substantially all theoretical predictions on hard diffraction 
and changes drastically the interpretation of experimental results in 
terms of the structure of the pomeron, as discussed below.
  
\subsection{Does the hard pomeron obey the momentum sum rule?}

We are now ready to answer the question: does 
the hard pomeron reported by UA8 obey the momentum sum rule?  
As mentioned above, the fact that the observed diffractive dijet rate 
is considerably smaller than the rates 
predicted for a pomeron with a fully hard-quark(gluon) 
structure function is generally interpreted as meaning 
that the momentum sum rule is violated.   
However, the predicted rates 
were calculated with the unnormalized DL flux factor
of Eq.~\ref{DL_flux}, whose integral over $t$ and $\xi$ within the 
range $0<|t|<\infty$ and $1\;GeV^2/s<\xi<0.1$ is 3.9 (using $\beta_0^2=
3.5$ GeV$^{-2}$ as in \cite{UA8-EPS}).  Therefore, 
with a normalized flux factor the predictions of \cite{UA8-EPS} 
for the dijet rates become 
3.9 times smaller.  This correction moves the UA8 
``discrepancy factors" of $0.46\pm0.08\pm0.24$  ($0.19\pm0.03\pm0.10$) 
for a hard-quark(gluon) dominated pomeron to the values 
$1.79\pm 0.31\pm 0.93$ ($0.74\pm 0.11\pm 0.39$), which are consistent 
with unity and therefore no longer in disagreement with the momentum sum rule.  

\subsection{Diffractive W's at the Tevatron and HERA physics}
Diffractive $W$ production probes the quark structure function of the pomeron. 
Using the flux factor of Eq.~\ref{BI_flux}, Bruni and Ingelman  
predicted \cite{BI} 
that the ratio of diffractive to non-diffractive $W$ production at
$\sqrt s$=1800 GeV is expected to be $\sim 17$\% (0.8\%, 0.3\%) for a 
hard-quark (hard-gluon, soft-gluon) pomeron structure function.   
With the flux factor of Eq.~\ref{R_flux}, which has a different $\xi$ 
and $t$ dependence, the predicted rate under the same kinematical conditions 
goes up to $\sim 24$\%.  However, using the flux scaling factor of 
$(1800/20)^{4\times 0.115}$ at $\sqrt s$=1800 GeV, brings the 
prediction down to $\sim 3\%$, 
the exact value depending somewhat on the parameters used
in Eq.~\ref{R_flux}. Therefore, in order to probe the pomeron for  
an {\em effective} $\sim 15\%$ hard-quark component in its
structure function,  the level predicted by 
Donnachie and Landshoff \cite{DL88}
on the basis of their analogy between the pomeron and the photon in the 
way they couple to quarks, the diffractive to non-diffractive 
$W$ production ratio must be measured with an accuracy smaller than 0.5\%.
 
Hard diffraction has also been under study in $e^-p$ collisions 
at $\sqrt s=$314 GeV at HERA, where 
virtual photons from 
30 GeV electrons collide with pomerons from 820 GeV protons.  
Diffractive events are identified 
by the rapidity gap method and the results are interpreted in terms of the 
structure function of the pomeron.  As discussed above, in drawing 
conclusions about the pomeron structure from the data by comparing 
them to theoretical predictions based on the Ingelman-Schlein model, 
the predictions must be reduced by the flux scaling factor, which 
at the typical $\gamma^{\star}p$ center of mass energy of $\sim 150$ GeV 
is $\sim 2.5$. 
 
\subsection{Events with a rapidity gap between two jets}

The exchange of a hard pomeron, which is a color-singlet or {\em colorless} 
QCD construct,  between a proton and an antiproton is 
expected to produce dijet events with a rapidity gap between the two jets. 
It was estimated \cite{Bj1} that the 
ratio, $R_{jets}$,  of the cross section for color-singlet exchange to
single gluon exchange
events with the same kinematics is $\approx 0.1<|S|^2>$, where $|S|^2$ is
the ``survival probability" for the gap, placed at 3--30\% \cite{Bj1,Bj2}.
According to this estimate, $R_{jets}$ 
should not depend strongly on the rapidity gap width. 
At the Tevatron collider at $\sqrt s= 1800$ GeV, 
the D0 collaboration reported an upper limit of 1.1\% (95\% CL) for such
events \cite{D0_gap}, and CDF reported a signal of $R_{jets}=
0.85\pm0.12(stat.)^{+0.0024}
_{-0.0012}(syst.)$ \cite{CDF_gap}, which is in agreement with the 
above prediction.  Typical rapidity gaps studied were around 2 units.

As can be seen from Table~\ref{Table_2}, the ratio of all events 
with a rapidity gap of two units 
to all the non-diffractive inelastic events at $\sqrt s=1800$ GeV 
is $1.25$\% (the first number in the Table divided by 
the non-diffractive inelastic cross 
section of 50 mb \cite{EDT}). This ratio, which we shall call $R_{soft}$, 
{\em is approximately 
the same as the ratio $R_{jets}$} measured by CDF!  We therefore 
propose that the exchange of a pomeron produces the same fraction of 
hard as soft interactions, which implies that the rapidity gap survival 
probability in dijet production is close to 100\%.  
A test for this model is provided by the rapidity gap width  
and energy dependence it predicts
for $R_{jets}$.  The dependence on the rapidity gap width at $\sqrt{s}=1800$ 
GeV is that of 
Table~\ref{Table_2}.  The dependence on energy is obtained by dividing 
the DD cross sections given in Table~\ref{Table_1}
by the corresponding inelastic non-diffractive cross sections, 
which can be obtained from the formulae given in this paper.
At the LHC, $\sqrt{s}=14$ TeV, we predict that 
$\sigma_T$=128.5 mb, $\sigma_{el}$=44.1 mb, $2\sigma_{SD}$=10.0 mb, 
$\sigma_{DD}=0.75$ mb, $\sigma_{DP}=52$ $\mu$b, $\sigma_{ND}$=73.6 mb, 
$\sigma_{DD}^{\Delta y=2}$=0.36, and therefore $R_{jets}=R_{soft}$=0.36/73.6
=0.5\% (for $\Delta y=2$),
which is less than one half of the prediction for the Tevatron.

\section{Conclusion}
Regge phenomenology, with simple pomeron exchange and factorization, 
describes well the general 
features of elastic, diffractive, and total cross sections. However, 
the energy dependence of the single diffraction cross section is not 
predicted correctly by the theory.  In fact, in a model of simple 
factorizable pomeron 
exchange with a constant triple-pomeron coupling, 
the diffractive cross section rises at a rate 
much larger than the total cross section, violating unitarity at 
the TeV energy scale.  
Nevertheless, the 
concept of factorization was extended  
to ``hard diffraction" 
processes in a model that assumes that a high energy proton carries 
along a ``pomeron flux" that interacts with {\em the other} proton 
producing jets, W's, or involving other high $p_T$ phenomena.  This model 
was employed by the UA8 Collaboration 
to interpret the results of an experiment 
designed to probe the structure of the pomeron in 
diffractive dijet production.  The UA8 results indicate that the pomeron has 
a hard partonic structure, but the reported ``hard pomeron" does not 
obey the momentum sum rule.  In this paper 
we show that by renormalizing the pomeron flux carried by a proton to unity, 
we obtain the correct energy dependence for the single diffraction cross 
section, and that with a renormalized pomeron flux the 
UA8 hard pomeron results become consistent with the momentum sum rule. 

In addition to single diffraction, we have calculated cross sections for 
double diffraction and double pomeron exchange and find agreement with 
available data.  Our results for double diffraction show a rapidity gap 
dependence that can be tested with accurate measurements at the Tevatron.  
Furthermore, noting that our prediction for the ratio of {\em soft} rapidity 
gap events 
(double-diffractive) to all non-diffractive events, $R_{soft}$, is close 
to the measured ratio for {\em hard} processes containing jets, $R_{jets}$, 
we propose that the observed dijets with a rapidity gap are due to 
the pomeron and that $R_{jets}=R_{soft}$. On the basis of this model,  
we then use our calculations for double diffraction to predict 
that the ratio $R_{jets}$ will decrease 
slowly with energy to become 0.5\% (for $\Delta y=2$) at the LHC.
  
Pomeron flux renormalization 
affects all predictions for hard diffraction processes made    
with an unnormalized flux, like those of Refs.~\cite{IS,BI}.  Such predictions 
must generally be reduced by the integral 
of the flux factor used in deriving them, exercising caution with 
regard to consistency between the parameters of the pomeron trajectory 
and the other parameters in the flux factor. 

\section{Acknowledgments}
I am indebted to my colleages 
P. Melese, A. Bhatti, S. Bagdasarov and A. Maghakian for many useful 
discussions,
and to A. Mueller for several critical 
comments.  This work has been supported 
in part by the Department of Energy under grant DE-FG02-91ER40651.

\newpage
\clearpage

\clearpage

\end{document}